# Find Central Dogma Again: Leveraging Multilingual Transfer in Large Language Models

Wang Liang

Huazhong University of Science and Technology, 430070, P.R. China

*To whom correspondence should be addressed. E-mail:wangliang.f@gmail.com

[Abstract]In recent years, large language models (LLMs) have achieved state-of-the-art results in various biological sequence analysis tasks, such as sequence classification, structure prediction, and function prediction. Similar to advancements in AI for other scientific fields, deeper research into biological LLMs has begun to focus on using these models to rediscover important existing biological laws or uncover entirely new patterns in biological sequences. This study leverages GPT-like LLMs to utilize language transfer capabilities to rediscover the genetic code rules of the central dogma. In our experimental design, we transformed the central dogma into a binary classification problem of aligning DNA sequences with protein sequences, where positive examples are matching DNA and protein sequences, and negative examples are non-matching pairs. We first trained a GPT-2 model from scratch using a dataset comprising protein sequences, DNA sequences, and sequences from languages such as English and Chinese. Subsequently, we fine-tuned the model using the natural language sentences similarity judgment dataset from PAWS-X. When tested on a dataset for DNA and protein sequence alignment judgment, the fine-tuned model achieved a classification accuracy of 81%. The study also analyzed factors contributing to this zero-shot capability, including model training stability and types of training data. This research demonstrates that LLMs can, through the transfer of natural language capabilities and solely relying on the analysis of sequences themselves, rediscover the central dogma without prior knowledge of it. This study bridges natural language and genetic language, opening a new door for AI-driven biological research.

## 1 introduction

Advancements in large language models have revolutionized the field of artificial intelligence, with profound implications for bioinformatics. Specifically, in nucleic acid analysis, specialized models such as DNABert2, HyenaDNA, and ScBert have emerged to address classification and structural prediction challenges related to DNA sequences (1-5). By leveraging the capabilities of large language models, these tools provide novel insights into genetic data. Concurrently, significant progress has been made in protein-related research, where models like ProTrans, ProteinBERT, and ESM2 excel in tasks such as predicting protein structures and annotating their functions (6-12). These innovations underscore the versatility of large language models in tackling complex biological questions, effectively bridging the gap between computational linguistics and molecular biology.

Many studies have focused on improving existing biological problems, more in-depth research on large models aims to leverage their reasoning and transfer capabilities to rediscover or assist in discovering important principles within the field. These studies are generally based on zero-shot methods or unsupervised learning.

For example, the Evo model uses protein, DNA, and RNA data to train a large language model based on the StripedHyena architecture, and it leverages zero-shot learning to predict the impact of mutations in different regions of gene sequences on the function of proteins, non-coding RNAs and regulatory DNA(13). In contrast, the GPN model employs genomic data to train a convolutional language model, which can accurately predict genome-wide variant effects in an unsupervised manner(14). megaDNA, a multiscale transformer model, is pre-trained on unannotated bacteriophage genomes with nucleotide-level tokenization, and it can predict gene essentiality across the phage genome in a zero-shot manner (15). The paper (16) demonstrates that by performing unsupervised pre-training on a large corpus of protein sequences, models such as ESM-1v are able to capture the structural and functional information within the sequences, thereby enabling direct prediction of mutation effects without any task-specific supervised data. Furthermore, DART-Eval has designed a standard evaluation dataset for large-scale gene language models, which includes metrics that can evaluate DNALMs in a zero-shot setting by calculating likelihoods and embeddings(17).

**Challenge**.Currently, most large biological sequence models that support zero-shot capabilities still rely on extracting sequence feature vectors for data analysis, and they are not yet able to leverage language model reasoning as effectively as ChatGPT. For research area like mathematics, although there is an abundance of formulaic language, these languages are interpretable, meaning they have clear natural language descriptions or definitions. Therefore, large models can be conveniently utilized through methods like prompt engineering. However, biological sequence languages remain largely unknown, significantly limiting the depth of large model applications in biological research. This makes it challenging to utilize the reasoning capabilities of large models to discover entirely new principles, as has been done in other disciplines. Although there are enormous differences between biological sequences and natural language, constructing a bridge that connects natural language with genetic language is key to solving the problem.

Research on language transfer capabilities within large models provides key insights into solving this problem.

An intriguing direction within these studies is the exploration of multilingual transfer capabilities in large language models. Research on multilingual transfer has demonstrated how large pre-trained models can effectively apply knowledge learned in one source language to other languages. These studies have validated the effectiveness of cross-linguistic knowledge transfer, including between different natural languages, as well as between programming languages and natural language (18-23). Moreover, Our previous research has shown that the transfer phenomenon from natural language capabilities to DNA sequences also exists, making it possible to directly apply natural language processing techniques to DNA sequence analysis (24).

Research(24) has demonstrated that fine-tuning models on English similar text pairs enables the assessment of similarity between two DNA or two protein sequences, achieving accuracies exceeding 80%. Building upon this foundation, we aim to further utilize large models to investigate the coding rules between DNA and protein sequences, specifically the relationship between these two modalities as described by the central dogma of molecular biology.

The central dogma—including the discovery of the genetic code—is the result of decades of accumulated experimental evidence. Marshall Nirenberg and Heinrich Matthaei, through experiments involving the artificial synthesis of RNA (such as poly-uridine UUU...) and cell extracts, discovered that UUU encodes phenylalanine, marking the first breakthrough in deciphering the genetic code. By 1966, a team of scientists (including Hal Gobind Khorana and others) had chemically synthesized specific sequences to decode all 64 codons, thereby confirming the universality of the genetic code.

In this paper, we aim to rediscover the key rules of the genetic code using only large models combined with biological sequence analysis—without any prior knowledge of the central dogma.

Specifically, we first reformulate the central dogma as a binary classification problem on DNA-protein sequence alignment. We then trained a GPT-2 model, named GPT2-gene-multi, from scratch using DNA sequences, protein sequences, and seven different natural languages. Subsequently, we fine-tuned this model using the English sentence similarity dataset from PAWS-X to obtain the fine-tuned model GPT2-gene-multi-ft. When fine-tuned model tested on datasets for DNA-protein sequence alignment, the classification accuracy can reach over 81%.

## 2 Materials and methods

### 2.1 Experiment Design

In current machine learning research, DNA sequences and their corresponding protein-coding sequences are often used as datasets to evaluate models' multimodal sequence processing capabilities. Typically, matching DNA-protein pairs are treated as positive samples, while non-matching pairs serve as negative samples. Supervised learning methods are then employed, where the model is trained on a portion of the dataset and tested on the remaining data.

In contrast, our study adopts an unsupervised approach, aiming to perform zero-shot predictions without any prior knowledge of the central dogma. Drawing inspiration from zero-shot experimental designs in natural language large models, we propose the following three unsupervised methods to predict the central dogma:

1 Designing a Binary Classification Task for DNA-Protein Sequence Alignment: This method mirrors the current DNA-protein coding determination datasets, using matching pairs as positive samples and non-matching pairs as negative samples. However, the dataset is solely utilized for testing purposes and is not incorporated into the training process. This approach leverages the generalization and transfer capabilities of pre-trained or fine-tuned models during testing.

2 Zero-Shot Similarity Calculation: Without any training, this method directly computes the cross-modal embedding similarity between DNA and protein sequences. By analyzing the similarity scores of positive and negative samples, a threshold is established to differentiate between them, thereby obtaining classification metrics such as accuracy. Additionally, this approach can further elucidate the similarity relationships between protein and DNA vocabularies.

3 Prompt-Based Zero-Shot Learning: This technique involves crafting natural language instructions to guide the model's reasoning process. For instance, a prompt might be:

*"You are a bioinformatics expert tasked with determining whether a given DNA sequence encodes the corresponding protein sequence.*

*DNA sequence: ATGGCGCTAAAG*

*Protein sequence: MART*

*Please answer 'yes' or 'no':"*

The model's generated response ('yes' or 'no') is then mapped to binary classification labels.

Among these methods, the first is the most commonly employed. In the context of causal language models (CLMs), the third method can often be viewed as a variation of the first. The second method, due to differences in tokenization and vector representations, poses challenges in determining similarity and is generally used for secondary validation.

Therefore, our study primarily utilizes the first method to design experiments aimed at discovering the genetic code rules within the central dogma.

## 2.2 Research Approach

A commonly used method for zero-shot prediction of DNA-protein datasets with large language models is leveraging linguistic transfer ability.

Multilingual transfer ability, which reflects how well models fine-tuned on one source language can be applied to other languages, has been extensively studied in multilingual pre-trained models. In this context, we explore the transfer of natural language abilities to genomic language.

The main process involves first pre-training a large language model, then fine-tuning it on a natural language dataset, and finally evaluating the fine-tuned model on a genomic sequence dataset. The key to achieving successful language transfer lies in identifying a natural language dataset that shares structural similarities with the genomic dataset.

We adopted GPT-2 as the foundation model structure. For pre-training, we used seven languages (including English and Chinese,etc) along with DNA and protein sequences, employing a unified BPE encoding. We then fine-tuned the model on the PAWS-X dataset, a natural language translation and paraphrase similarity judgment dataset, due to its structural similarity to the DNA-protein alignment dataset. Finally, we evaluated the fine-tuned model on the DNA-protein alignment dataset.

The training and test process is illustrated in (Fig. 1).

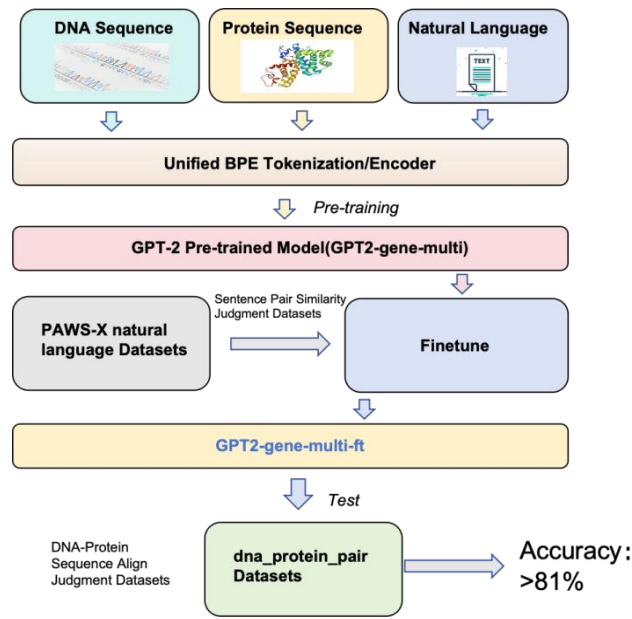

Fig.1. Training and Testing Process of the Research Approach. This model utilizes genomic and protein sequence data, along with encyclopedic text in multiple languages (including English and Chinese,etc), as pre-training corpora. The BPE (Byte Pair Encoding) method is employed for uniform tokenization, treating all text types equivalently without differentiation.The model is trained from scratch based on the GPT-2 small architecture, resulting in the pre-trained model GPT2-gene-multi. It is then fine-tuned on the PAWS-X similarity classification dataset, producing the GPT2-gene-multi-ft model. Finally, this fine-tuned model is evaluated on a DNA-protein alignment classification dataset, achieving a test accuracy of 81%.

A crucial aspect of achieving effective language transfer is identifying natural language datasets that share structural similarities with DNA-protein encoding. Several types of datasets can serve as potential candidates:

1. Machine Translation – The DNA → protein encoding process is analogous to translating a source language into a target language.

2. Semantic Similarity Texts – Similar to machine translation, this includes sentence pairs with equivalent meanings, such as two semantically similar English sentences or an English sentence paired with its semantically equivalent French translation.

3. Speech-to-Text – The process of encoding DNA into proteins resembles converting speech signals into recognized text.

4. Code Generation – DNA-to-protein conversion follows strict rule-based mappings, similar to the transformation between natural language and programming code.

Given the availability and diversity of datasets, we chose semantic similarity text datasets as candidates for language transfer learning. One of the most representative datasets in this category is PAWS-X, a cross-lingual sentence pair classification dataset designed to evaluate and improve NLP models' ability to recognize semantic equivalence across languages.

Each PAWS-X sample typically includes the following fields:

- sentence1: The first sentence.

- sentence2: The second sentence.
- label: Indicates whether the two sentences are synonymous (0 = not synonymous; 1 = synonymous).
- lang: The language code of the sample.

The sentence pairs in PAWS-X can be in the same language or different languages, making it structurally similar to translation datasets.

Our approach involves fine-tuning the model on the PAWS-X dataset, learning the sentence1 → sentence2 mapping. We then evaluate the fine-tuned model on a DNA → protein alignment dataset. If the model achieves promising results, it suggests that the model has successfully acquired the ability for zero-shot discovery of the genetic code.

## 2.3 Training datasets

**Pretrained datasets**

The training data is roughly outlined in the (Table.1):

Table.1 Training data

| Data type | source | data size |
|---|---|---|
| DNA sequence | multiple model organism genomes | 4G |
| Protein sequence | Swiss-Prot/TrEMBL | 4G |
| Natural Language text | wiki. en,fr,zh,de,es,ja,ko | 4G for each language |

The pretraining datasets include a diverse range of data sources. The training data consists of DNA sequences from multiple model organism genomes (4GB), protein sequences from Swiss-Prot/TrEMBL (4GB), and natural language text from Wikipedia in multiple languages (English, French, Chinese, German, Spanish, and Japanese), with 4GB allocated for each language. This dataset composition ensures a comprehensive foundation for training the model on both biological and linguistic data.

**PAWS-X Dataset**

PAWS-X is a sentence-pair similarity assessment task where each data entry consists of two sentences. A typical data entry looks like this:

{

 'sentence1': 'In Paris, in October 1560, he secretly met the English ambassador, Nicolas Throckmorton, asking him for a passport to return to England through Scotland.',

'sentence2': 'In October 1560, he secretly met with the English ambassador, Nicolas Throckmorton, in Paris, and asked him for a passport to return to Scotland through England.',

 'label': 0

}

The task is to determine whether the two sentences express the same meaning, which is a binary classification problem. To extend this to other languages, the English sentence pairs were translated into six target languages (French, Spanish, German, Chinese, Japanese, and Korean) by professional translators and manually checked to ensure quality and consistency.

We fine-tune the model by selecting same or different language combinations for sentence1 and sentence2, and then evaluate it on the DNA-protein alignment dataset.

**DNA-Protein Pair Align Judgment Task**

Based on our experimental design and referring to the PAWS-X construction method, we built a DNA-protein pair alignment judgment task based on gene coding rules. If a protein sequence corresponds to its DNA coding sequence, it is considered a positive example; otherwise, it is a negative example. The specific data design was inspired by the dataset used in Lucaone's paper (25). A typical data entry is structured as follows:

```
{
    "sentence1": "ACCAGTGCTCAGGTTAACAAAATAATAAAAGGAA....",
    "sentence2": "MASGRLQLLAFALALSGVSGVLAATLLPNWTVSVD...",
    "label": 0
}
```

The Lucaone's dataset contains approximately 25,600 entries, with a balanced number of positive and negative examples.

In Lucaone's DNA-Protein dataset, to increase the testing difficulty, each DNA sequence was extended by adding 100 extra nucleotides at both the 5' and 3' ends, thereby preserving intron sequences within the data. In our version, we have removed these additional 100 base pairs (bp).

Since this dataset is used only for testing, we selected 4,000 samples from it. This dataset is referred to as dna_protein_pair.

In the Lucaone's paper, part of negative examples were generated by keeping the DNA sequence unchanged and modifying the corresponding protein sequence through data perturbation methods, such as:

- Amino Acid Substitution (translation errors)
- Shuffling Protein Sequence

To further reduce the test difficulty, we constructed negative examples using random protein sequences. Specifically, we kept the DNA sequence unchanged, used the correct protein-coding sequence as a reference, and then randomly selected a protein sequence of similar length from a protein database, ensuring its sequence similarity to the reference was below a threshold (e.g., 0.45). This dataset is referred to as dna_protein_pair_rand.

The complete construction method can be found in Section 1 of the supplementary material.

The fine-tuning and testing datasets are detailed in (Table.2).

Table.2 Finetune and test data

| dataset name | Data type | source |
| --- | --- | --- |
| PAWS-X | Natural Language Pair | PAWS-X |
| dna_protein_pair | DNA-protein pair | NCBI-Refseq |
| dna_protein_pair_rand | DNA-protein pair | Swiss-Prot/TrEMBL/NCBI |

## 2.4 Training Strategy

**Model Structure**

The base model uses the exact same model architecture as GPT-2, with the primary difference being the training corpus. For the research validation, the base model is structured as GPT-2 Small, which can be trained on a single 4090 GPU. For classification fine-tuning, a typical classification head is used, outputting two classes. For prompt instruction fine-tuning, the same head as the petrained model is used, and the training mode is also identical to that of the pre-training. The model structure is shown in (Fig.2).

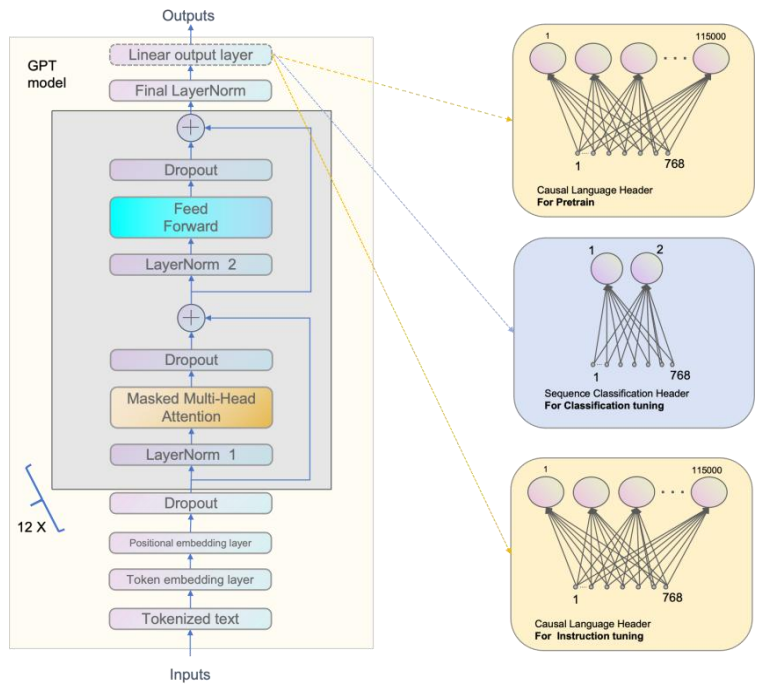

Fig.2. Model structure. The model architecture is identical to that of GPT-2 small. Pre-training and instruction tuning use the same causal language header,the layer mapped 768 hidden units to 115,000 units (the number of tokens in the vocabulary). Classification uses sequence classification header, The layer maps from 768 hidden units to only 2 units, where the 2 units represent the two classes id.

### Tokenization

Research on language capability transfer indicates that using a unified tokenizer is a necessary condition for transferring capabilities. Both the Byte Pair Encoding (BPE) tokenizer used in GPT2 models and the WordPiece tokenizer used in BERT models can process DNA and protein sequences. These tokenizers typically split sequences into tokens of 1 to 6 characters, which is a common method for feature extraction in biological sequences. Therefore, the choice of tokenizer does not affect our validation of language capability transfer. Here, we use the default BPE tokenizer of the GPT-2 model.

### Pretrain

According to the typical design of GPT-2, it accepts sequences with a maximum length of 1024 as input. We used the same architecture as the GPT-2 Small model, which consists of 12 Transformer layers, each with 768 hidden units and 12 attention heads. This small model has approximately 117 million parameters.We trained the GPT-2 model using mixed-precision floating-point arithmetic on a machine equipped with a single Nvidia 4090 GPU. We employed a dynamic learning rate schedule, and the model was trained for a total of 3 to 5 epochs.

### Finetune

We used the GPT2-small model as a baseline for fine-tuning on the PAWS-X English dataset. Subsequently, we evaluated the model on DNA-protein pair align judgment datasets.

Input and Output Format:

we concatenated the two sequences directly as the input string, with the label serving as the classification ID:

Input: sequence1 + sequence2

Output: label

Note: For GPT-based models, no separator is needed as the model itself determines the boundaries between the two sequences.

Fine-tuning Procedure:

For Classification Fine-tuning,The model's header is set to the commonly used softmax classification form. We utilized the same training tricks across all the applications, where the learning rate was first linear warmed-up to the peak value and then linear decayed to near 0.

Due to the smaller scale of model parameters and the datasets size being in the tens of thousands of samples, we used the full-parameter fine-tuning method. Fine-tuning can be completed within about an hour on a single 4090 GPU.

**Model Evaluation**

We fine-tuned the model using the PAWS-X dataset and tested it on the DNA-protein alignment classification dataset. The model's accuracy is used to measure its performance. Since this is a binary classification task, the meaning of classification IDs 0 and 1 may differ during transfer. Therefore, if the test accuracy is below 0.5, we can simply swap label IDs 0 and 1. For some experiments, we further evaluate the model using a Confusion Matrix.

## 3 Experimental Results

### 3.1 Basic results

As a binary classification task, we primarily used accuracy as the evaluation metric.

We utilized three pretrained models as the base models. Among them, GPT2 is the standard model released by OpenAI, primarily trained on English data. The gpt2-gene-eng model is a GPT2 model that we trained from scratch using a dataset consisting of English text, DNA sequences, and protein sequences. The gpt2-gene-multi model is also a GPT2 model trained from scratch, but its training data includes DNA and protein sequences, as well as text data in seven natural languages, including English and Chinese.

These base models were fine-tuned using the English-English sentences similarity classification corpus from PAWS-X and then tested on the DNA-protein alignment classification dataset. In alignment dataset, we inverted the original labels, meaning that 0 represents alignment, and 1 represents non-alignment.

As a comparison, we also evaluated the fine-tuned models on the PAWS-X English test set, as well as the French and Chinese test sets. The specific test results are shown in (Table.3).

Table.3 Accuracy of different pretrained models in different datasets

| pretrained model | pretrain dataset | finetune dataset | test-en | test-fr | test-zh | test dna_protein_pair | test dna_protein_pair_rand |
|---|---|---|---|---|---|---|---|
| gpt2 | en | en-en | 0.90 | 0.77 | 0.63 | 0.57 | 0.55 |
| gpt2-gene-eng | en, dna, protein | en-en | 0.89 | 0.74 | 0.65 | 0.59 | 0.68 |
| gpt2-gene-eng_zh | en, zh, dna, protein | en-en | 0.89 | 0.73 | 0.72 | 0.67 | 0.63 |
| gpt2-gene-eng_zh_de_es | en, zh, de, es, dna, protein | en-en | 0.89 | 0.75 | 0.72 | 0.66 | 0.63 |
| gpt2-gene-mulit | dna, protein, Multi-lan | en-en | 0.90 | 0.82 | 0.74 | 0.73 | 0.76 |
| gpt2-gene-eng | en, dna, protein | dna_protein_pair | - | - | - | 0.94 | 0.54 |
| gpt2-gene-eng | en, dna, protein | dna_protein_pair_rand | - | - | - | 0.54 | 0.99 |

The comparison of base models highlights the impact of pretraining data on cross-lingual and cross-modal transfer learning. The standard GPT-2 model, pretrained only on English text, achieved 0.90 accuracy on PAWS-X English, but performed poorly on French (0.77) and Chinese (0.63). Its accuracy on DNA-protein pair classification was 0.57, with a slight drop to 0.55 on randomly paired sequences. The GPT2-Gene-Eng model, pretrained on English, DNA sequences, and protein sequences, slightly underperformed in PAWS-X English (0.89) but showed minor improvements in DNA-protein classification (0.59) and a notable boost on random protein pairs (0.68).

GPT2-gene-eng_zh and GPT2-gene-eng_zh_de_es, two models trained on DNA sequences, protein sequences, and a subset of natural language sequences, exhibit language transfer performance that lies between that of GPT2-gene-eng, which is trained solely on DNA and protein sequences, and GPT2-gene-multi, which is trained on the complete set of sequence data.

The GPT2-Gene-Multimodel, pretrained on DNA, protein sequences, and seven natural languages, maintained high PAWS-X English accuracy (0.90), significantly improved French (0.82) and Chinese (0.74) performance, and demonstrated the best performance in DNA-protein classification (>0.7 for both datasets), indicating strong adaptability across both natural and biological languages.

For comparison, we also tested the classification performance of supervised learning by using a portion of dna_protein_pair or dna_protein_pair_rand as training data for model fine-tuning, while reserving another portion for testing and validation. The results show that supervised learning achieves a classification accuracy of over 0.94, significantly outperforming the zero-shot prediction method based on language transfer ability. This result is also in line with expectations.

The best transfer performance was achieved by the GPT2-Gene-Multi model on the dna_protein_pair_rand dataset. The Confusion Matrix for this test result is shown in the (Fig.3).

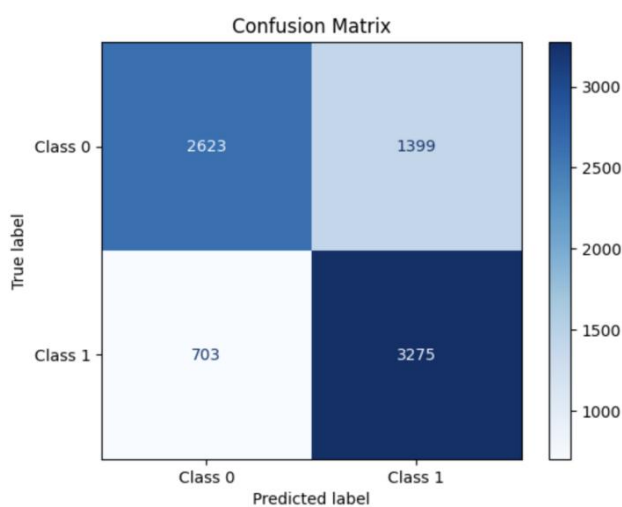

Fig.3 Confusion Matrix of DNA-Protein Alignment Classification. After fine-tuning the gpt2-gene-mulit model on the PAWS-X English similarity dataset, testing on the dna_protein_pair_rand dataset achieved a classification accuracy of approximately 76%. In this evaluation, class 0 represents aligned data and class 1 represents non-aligned data. While the positive and negative samples are relatively balanced, a higher proportion of class 0 samples were misclassified as class 1. Overall, the classification performance is essentially normal.

Based on the above data analysis, we can draw the following conclusions:

- Cross-lingual Pretraining Improves Generalization:The GPT2-Gene-Multi model, which included multilingual data, outperformed both GPT-2 and GPT2-Gene-Eng in non-English PAWS-X tasks and DNA-protein classification.

- Adding DNA and Protein Data Helps with Biological Tasks:The GPT2-Gene-Eng and GPT2-Gene-Multi models both performed better than the standard GPT-2 on DNA-protein alignment tasks.

- Multilingual Pretraining Offers the Best Performance Overall: The GPT2-Gene-Multi model achieved the highest accuracy across all test datasets, confirming the effectiveness of incorporating both biological and multilingual data.

### 3.3 Training Stability

In the training of large models such as GPT-2, the choice of seed value can introduce variability in training outcomes (18). Our experiments demonstrated that when transferring English language capabilities to other natural languages, the influence of different seed settings is relatively minor. In contrast, when transferring these capabilities to DNA-Protein Sequences, the impact of seed settings is considerably more significant.

To assess the impact of seed settings, we randomly assigned different seed values and repeated the experiment multiple times, collecting statistics on the fine-tuning and testing results.

As a comparison, we trained two base pre-trained models, GPT2-gene-multi-v1 and GPT2-gene-multi-v2, both trained on DNA and protein sequences as well as seven natural languages, with the only difference being the random seed settings. Based on these two pre-trained models, we conducted multiple fine-tuning experiments and evaluated the accuracy results. The statistical outcomes are illustrated in the (Table.4):

Table.4 Accuracy of differet pre-trained model

| model | finetune | test en pairs | test fr pairs | test zh pairs | test dna_protein_pair | test dna_protein_pair_rand |
|---|---|---|---|---|---|---|
| gpt2-gene-mulit-v1 | en-en | 0.90 | 0.82 | 0.73 | 0.73 | 0.71 |
| gpt2-gene-mulit-v2 | en-en | 0.90 | 0.82 | 0.74 | 0.67 | 0.76 |

The two pre-trained models, GPT2-gene-multi-v1 and GPT2-gene-multi-v2, were fine-tuned on the English pairs similarity judgment dataset of PAWS-X and then tested on the PAWS-X datasets in English pairs, French pairs, and Chinese pairs, as well as on the dna_protein_pair and dna_protein_pair_rand datasets.

It can be observed that models pretrained with different seeds show minimal variation (within 1%) in test results for natural languages such as English and French. However, when tested on DNA-protein sequences, the accuracy difference exceeds 5%.

During fine-tuning, the choice of seed also has a significant impact on performance. We conducted multiple fine-tuning experiments using different seeds and tested the models on the DNA-protein alignment dataset. The pre-trained model used is gpt2-gene-multi-v2, and the fine-tuning dataset consists of English similarity pairs. After fine-tuning, the model was tested separately on the dna_protein_pair and dna_protein_pair_rand datasets. The accuracy distribution is shown in the (Fig.4).For comparison, the test accuracy on French and Chinese datasets is also included.

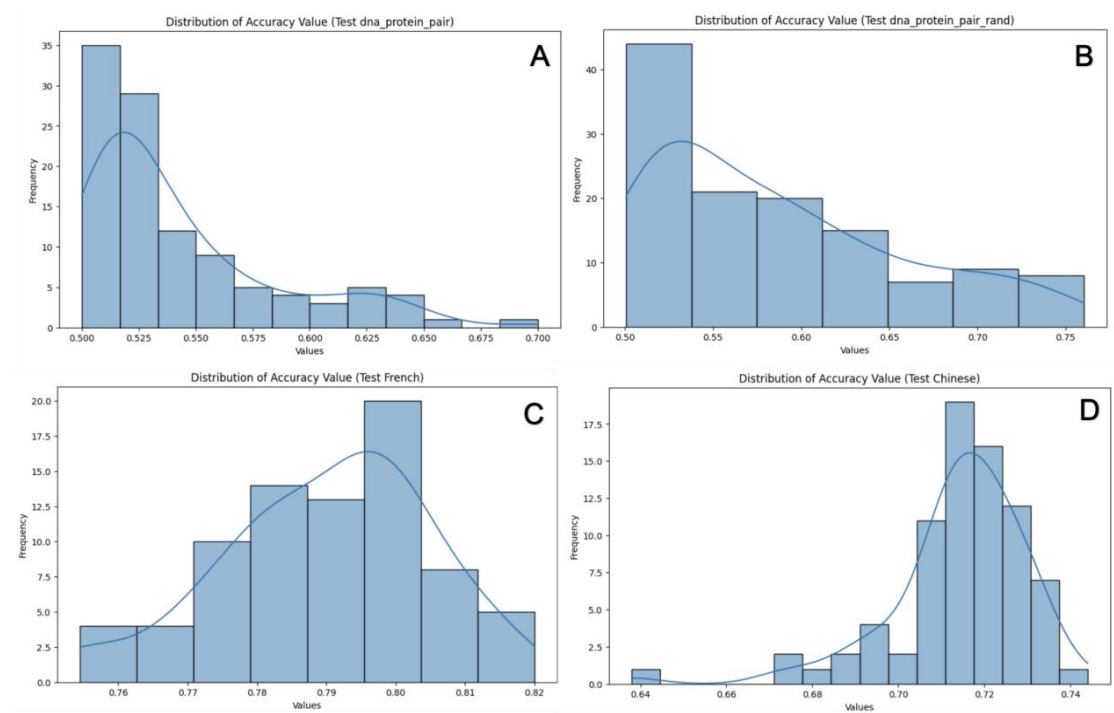

Fig.4 Accuracy Distribution Using Different seeds. The pre-trained model is gpt2-gene-multi-v2, and the fine-tuning data consists of English sentence pairs from PAWS-X. During the analysis of the fine-tuned model, different random seeds were used to evaluate their impact on the test accuracy results. Figure A and Figure B show the accuracy distribution on the DNA-protein alignment dataset:A corresponds to the dna_protein_pair dataset. B corresponds to the dna_protein_pair_rand dataset.

If we set 70% accuracy as a threshold for good performance:In Figure A, only 1% of the results exceed 70%. In Figure B, only 9% of the results exceed 70%. This implies that for the dna_protein_pair dataset, more than 100 fine-tuning trials are needed to obtain a well-performing model. Most results are only slightly better than random guessing, and the difference between the highest and lowest accuracy values exceeds 0.2.

Figure C and Figure D show the test accuracy distribution on French pairs and Chinese pairs datasets, respectively. Except for a single outlier (0.64) in Figure D, the accuracy variation between the highest and lowest values remains within 0.08.

The data in the figure illustrates that when transferring language capabilities from English to gene language (DNA and protein sequences), the solution space of the optimization problem is highly rugged. Only a few random seeds enable the model to find local optima that perform well on both English data and gene sequences. The detailed analysis is as follows:

1. Non-convexity of the Solution Space and Scarcity of Local Optima

During fine-tuning, we are effectively searching for the optimal parameters in a high-dimensional non-convex loss space. Since the source language (English) and the target language (gene language) differ significantly in structure, symbol systems, and syntactic rules, most local minima are only effective for English data and do not generalize well to gene sequences. Only when a random seed happens to initialize the model and guide the optimization process towards those rare and "cross-domain-friendly" local minima, effective language transfer can be achieved.

2. Challenges in Aligning Cross-Domain Representations

As a natural language, English has linguistic structures, vocabulary, and syntactic rules that are completely different from the encoding rules of DNA or protein sequences. Successful transfer requires the model to restructure or reorganize some of its internal representations during fine-tuning so that it can both retain the linguistic capabilities learned from pretraining on English and capture the encoding patterns of gene sequences. The fact that only a few random seeds achieve this balance suggests that in most cases, the model's internal representation tends to solidify around source language features rather than naturally aligning with gene language.

3. High Variance and Instability

The observation that only a few seeds lead to good performance also highlights the high variance in cross-domain fine-tuning. Due to random initialization, data ordering, and other stochastic factors, the optimization trajectory can differ significantly across runs. This instability suggests that the current approach to cross-language (especially cross-domain) transfer is highly sensitive and may require additional strategies—such as model ensembling, weight averaging, or more robust optimization techniques—to reduce the impact of random seeds and improve the stability of the transfer.

4. Potential Approaches to Address Randomness

- Multiple Trials and Ensembling: Since only a small fraction of random seeds can locate ideal cross-domain minima, multiple fine-tuning experiments can be conducted, followed by model ensembling or weight averaging, to reduce the influence of individual unstable solutions.

- Regularization and Domain Adaptation: Introducing regularization techniques (e.g., dropout, gradient clipping) or domain adaptation methods can help the model smoothly bridge the distribution gap between the source (English) and target (gene language) domains.

- Mixed Training Strategy: During fine-tuning, training on a mixture of English and gene sequence data may help the model align representations across different languages, reducing its dependency on random seeds. However, this approach would shift the task from zero-shot transfer to few-shot learning.

Our approach of conducting multiple fine-tuning runs with different random seeds partially mitigates the challenges of language transfer, but it comes with a significant computational cost. Future research can leverage the insights from this analysis to reduce computational overhead and achieve faster and more stable convergence towards optimal solutions.

## 3.3 Best finetune Language

The PAWS-X test dataset includes not only English similarity text pairs but also similar text pairs in other languages such as French, German, and Chinese. We conducted further testing across all these languages.

The pretrained model used in our experiments was GPT2-Gene-Multi-v2, which was trained on protein sequences, DNA sequences, and all natural language data from PAWS-X. Specifically, we fine-tuned models using different natural language similarity sequences and then tested them on the DNA-protein alignment dataset. As a comparison, we also evaluated the fine-tuned models on English, French, and Chinese test sets.

Since fine-tuning performance is highly sensitive to random seed selection, we conducted 100 training runs for each dataset, with random seeds ranging from 1 to 100, and then selected the best-performing model for comparison (Table.5).

Table.5. Effect of Fine-Tuning on same language pairs datasets. Accuracy results.

| finetune | test-en pair | fr pair | zh pair | dna_protein_pair | dna_protein_pair_rand |
|---|---|---|---|---|---|
| en-en | 0.90 | 0.82 | 0.74 | 0.71 | 0.77 |
| fr-fr | 0.85 | 0.85 | 0.74 | 0.68 | 0.73 |
| de-de | 0.82 | 0.80 | 0.73 | 0.74 | 0.78 |
| es-es | 0.85 | 0.83 | 0.72 | 0.72 | 0.76 |
| zh-zh | 0.72 | 0.74 | 0.77 | 0.73 | 0.67 |
| ja-ka | 0.75 | 0.73 | 0.76 | 0.72 | 0.68 |
| ko-ko | 0.70 | 0.71 | 0.73 | 0.74 | 0.64 |

It can be observed that the prediction results can be clearly divided into two groups based on the choice of fine-tuning language. One group consists of Latin-based alphabetic languages, including English, French, German, and Spanish, which exhibit similar transfer effects and perform better on the dna_protein_pair_rand dataset. The other group includes East Asian languages such as Chinese, Japanese, and Korean, which also show comparable performance among themselves and achieve better results on the dna_protein_pair dataset. Relatively speaking, German demonstrates the best prediction performance, although its advantage is not particularly significant.

We also tested various combinations of different languages, such as English-French, Chinese-German, etc. The better-performing fine-tuning data combinations are listed in the table below (Table 6):

Table.6. Effect of Fine-Tuning on Different Language pair Datasets. Accuracy results.

| finetune | test dna_protein_pair | test dna_protein_pair_rand |
|---|---|---|
| en-fr | 0.69 | 0.78 |
| fr-es | 0.70 | 0.76 |
| de-zh | 0.74 | 0.76 |
| es-de | 0.71 | 0.79 |
| zh-ko | 0.68 | 0.81 |
| ja-es | 0.68 | 0.79 |
| ko-fr | 0.73 | 0.72 |

It can be observed that the fine-tuning performance using combinations of seven natural languages is not significantly different from that using same-language data. Among these, the most effective dataset is the zh-ko (Chinese-Korean) sentence pairs, achieving an accuracy of 0.81 on the dna_protein_pair_rand dataset, which is the best test result reported in this study. Detailed results of all tests are provided in Section 3 of the supplementary material.

Overall, the impact of random seeds is greater than the choice of source language. If a more stable and robust fine-tuning method can be found, the comparison of source language transfer effects would be clearer.

### 3.4 Pre-trained Model Prediction without Finetune

The choice of fine-tuning and pre-training seeds has a significant impact on the final classification results. This gives us an insight: if we set different random seeds for the pre-trained model without fine-tuning, can it still classify DNA-protein alignment data? We adopted a similar approach as in the previous section, using random seeds from 1 to 100, and directly predicting the DNA-protein alignment data with the pre-trained model without fine-tuning.

If the results are good, it would suggest that the pre-trained model has learned part of the central dogma and can zero-shot directly predict the DNA-protein encoding rules. If the performance is better than that of a fine-tuned model based on natural language, it would imply that the previous fine-tuning tests only demonstrated the zero-shot ability of the pre-trained model, rather than the transfer ability of language ability. Another possibility is that the pre-trained model shows some effectiveness, but slightly worse than the fine-tuned model, which would suggest that the fine-tuned model's predictive ability is a combination of the pre-trained model's zero-shot ability and the fine-tuned model's language transfer ability.

We used the original GPT-2 model, GPT-2-Gene-Eng, and GPT-2-Gene-Mulit pre-trained models to directly predict DNA-protein alignment data, taking the best result. For comparison, we also trained a model using only DNA and protein sequence data, GPT-2-Gene. The precision of predictions from each pre-trained model is shown in the Table.7 below:

Table.7. Accuracy results of pre-trained models directly predicting DNA-protein sequences.

| pretrained model | Pretrain dataset | test dna_protein_pair | test dna_protein_pair_rand |
|---|---|---|---|
| gpt2 | en | 0.55 | 0.53 |
| gpt2-gene | dna,protein | 0.63 | 0.58 |
| gpt2-gene-eng | dna，protein,en | 0.67 | 0.69 |
| gpt2-gene-mulit | dna，protein，Multi-lan | 0.69 | 0.68 |

As we can see, the GPT-2 model, which has not learned any DNA or protein sequences, performs similarly to random prediction. However, GPT-2-Gene, after learning DNA and protein sequences, achieves a prediction accuracy of 0.63 even without fine-tuning, which is significantly better than random prediction. Additionally, when it continues to learn natural language, its prediction accuracy improves slightly, suggesting that some patterns of genetic language might also be embedded within natural language sequences.

The best accuracy achieved by an no-finetuned model in direct prediction is 0.69, while the fine-tuned models generally achieve prediction accuracy around 0.75, with the best surpassing 0.81. This indicates that fine-tuning can further enhance the prediction accuracy for DNA-protein alignment, which shares similar structures. To make the results clearer, we compare the prediction accuracy of the four models, both before and after fine-tuning, in one unified Fig.5:

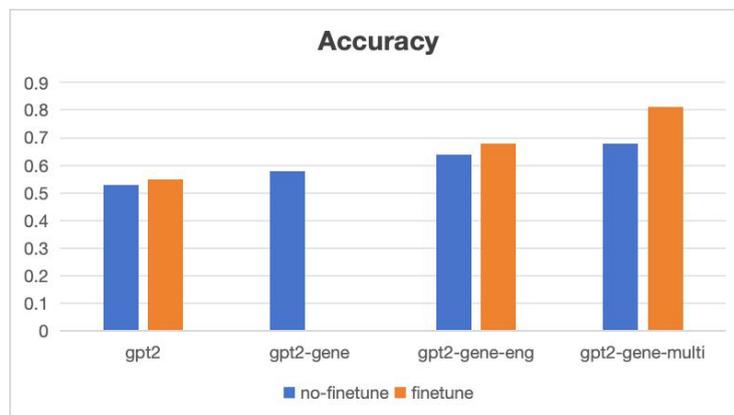

Fig.5. Prediction accuracy of different models on DNA-protein alignment data. The blue represents predictions made by the pre-trained model without fine-tuning, while the orange represents the results from the model fine-tuned with PAWS-X natural language similarity data.

From the above figure, we can observe two main trends. First, the more language types the model learns, the better its prediction results, whether it's direct prediction or prediction after fine-tuning. Second, fine-tuned models generally perform better than pre-trained models with direct prediction.

Regarding the results, we can analyze them from the model perspective. For large language models performing classification tasks, the base pre-trained model typically uses a softmax-based classification header. This layer takes the embedded vectors generated by the pre-trained model as input and outputs the classification result. The parameters of this layer are randomly initialized before fine-tuning. The reason why predictions still perform well without fine-tuning, meaning the classification header is completely random, is that the sequence vectors output by the pre-trained model already have a clear boundary between the two classes. As a result, even with a random classification header, there is a certain probability of achieving good classification results. We can use PCA (Principal Component Analysis) for dimensionality reduction to display the vector distribution of positive and negative examples in a 2D plane, as shown in the Fig.6 below.

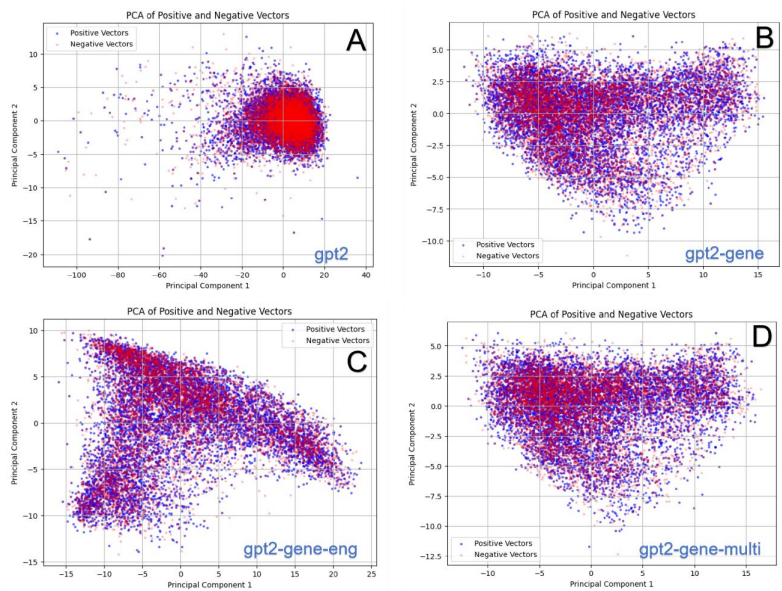

Fig.6. PCA 2D visualization of positive and negative example vectors, where blue represents positive examples and red represents negative examples. In panel A, the vectors are generated by the GPT-2 model. In panel B, the vectors are generated by the GPT-2-Gene model. In panel C, the vectors are generated by the GPT-2-Gene-Eng model. In panel D, the vectors are generated by the GPT-2-Gene-Multi model

As we can see, in the GPT-2 model, most of the data points in the two groups of vectors are tightly mixed together. However, after learning DNA and protein sequences, the GPT-2-Gene model generates two groups of vectors that are clearly more dispersed. After further learning natural language knowledge, the GPT-2-Gene-Eng model shows an even greater trend of dispersion in the two groups of vectors. This indicates that both the DNA and protein sequence knowledge learned during pre-training, as well as the natural language knowledge, contribute to the classification accuracy.

We can also use nearest neighbor distance to measure the distance between the two groups of vectors, which involves finding the nearest negative example vector for each positive example vector, calculating the distance, and then obtaining the average of the nearest neighbor distances for all positive examples. The larger the nearest neighbor distance, the clearer the distinction. The specific calculation results are shown in the Fig.7 below.

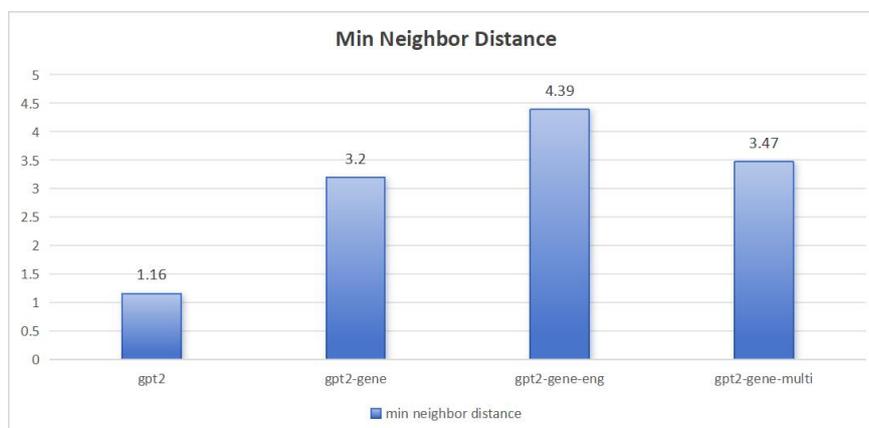

Fig.7. Average distance between positive and negative example vectors. The vectors are generated by four pre-trained models: GPT-2, GPT-2-Gene, GPT-2-Gene-Eng, and GPT-2-Gene-Multi. Each vector corresponds to a DNA-protein sequence pair. For the vectors generated by the GPT-2 model, the average distance between positive and negative examples is only 1.16, indicating poor distinguish ability. After learning DNA and protein sequences, the GPT-2-Gene model increases the average distance between positive and negative example vectors to 3.2. When further trained on English data, the GPT-2-Gene-Eng model increases the average distance to 4.39. However, after continuing to learn more languages, the GPT-2-Gene-Multi model shows a slight decrease in the average distance between positive and negative example vectors, dropping to 3.47.

## 4 Discussion

### 4.1 Necessary Conditions for Transfer Ability

These experimental results indicate that there are several key necessary conditions for successfully transferring natural language capabilities to genetic language:

**1. Multimodal Data Training**

During the pretraining phase, using a diverse set of training data—including multiple natural languages (not just English) as well as DNA and protein sequences—enables the model to learn a more comprehensive representation across different languages and symbolic systems. This approach allows the model not only to grasp English grammar and expression patterns but also to capture structural and sequential patterns in biological sequences. Due to the diverse data sources, the model develops a richer internal representation, making it more capable of adapting to genetic language compared to a model pretrained solely on English. This is similar to how a multilingual person can more easily find commonalities when encountering an unfamiliar field compared to a monolingual individual.

**2. Unified Tokenization and Encoding Methods (e.g., BPE)**

Using a unified tokenization (or encoding) method ensures consistency when converting different types of data (natural language and genetic sequences) into model inputs. For example, Byte Pair Encoding (BPE) can segment both English text and DNA/protein sequences into subwords or sub-sequences, forming a shared vocabulary or encoding space. This uniformity reduces discrepancies caused by inconsistent encoding, allowing the model to transfer knowledge more naturally and improving performance in cross-domain tasks. Many studies have also validated that unified encoding plays a crucial role in cross-lingual or cross-domain transfer learning.

**3. Fine-tuning on Natural Language Datasets with Structural Similarity to Genetic Sequence Tasks**

The structure of the dataset used for fine-tuning is critical for transfer effectiveness. Genetic sequence tasks often follow fixed patterns, such as DNA-protein alignment problems. Selecting a natural language dataset with a similar structure for fine-tuning allows the model to learn corresponding relationships and patterns, making it easier to transfer this "mapping" ability to genetic tasks. In other words, if the fine-tuning dataset shares formal and logical similarities with the target task, the model can more effectively find an appropriate solution when adapting to the new task.

These three conditions complement each other:

- Multimodal pretraining equips the model with a broad knowledge base.
- A unified tokenization method establishes a consistent internal representation space.
- Fine-tuning on structurally similar datasets helps the model effectively apply its learned general capabilities to domain-specific tasks (such as genetic language).

Together, these factors enable a more robust and efficient transfer of natural language abilities to genetic language processing.

## 4.2 The Randomness of Transfer Ability

From our experiments, we observed that randomness is one of the main characteristics of transferring natural language capabilities to genetic language. This phenomenon can be compared to an "accidental alignment of thinking patterns."

Imagine a model pretrained on a large amount of English data—similar to how a native English speaker is accustomed to English logic and expression patterns. When processing information, the model internally builds patterns that align with English language structures and thought processes.

However, while DNA and protein sequences are also "sequences," they have their own unique "grammar" and encoding rules, which are completely different from English language rules. In most cases, the model's "English thinking style" does not effectively apply to decoding biological sequences, leading to poor performance on the target task (DNA/protein alignment classification).

Occasionally, though, due to differences in random seeds, the initialization and data order during training might lead the model to "accidentally" discover a special local minimum—a point that is not only effective for English data but also happens to capture certain key encoding patterns in biological sequences. This is analogous to an English thinker suddenly realizing that their problem-solving habit also happens to work for deciphering a certain type of code, resulting in unexpectedly good performance.

Implications of This Phenomenon:

1. Cross-domain transfer is highly sensitive: Transferring a model between English and biological sequences requires finding those rare solutions that can accommodate both language structures. However, such solutions are extremely scarce in the optimization space and only appear under a few specific random seeds.

2. The alignment of thinking patterns is accidental: The model's English-based reasoning only produces good transfer results when it "happens" to overlap with the patterns of DNA or protein sequences under certain conditions. This suggests that cross-domain transfer may require additional regularization or domain adaptation techniques to stabilize such accidental alignments, making the results less dependent on random initialization.

Essentially, most of the time, an English thinker would struggle to decode biological sequences effectively. However, in rare cases, their way of thinking might happen to align with certain patterns in biological sequences, enabling them to "crack the code" successfully. While this occurrence is rare, it also demonstrates that the model possesses potential cross-domain transferability, though it remains fragile and highly random.

## 4.3 New Research Methods

After verifying that the model possesses the ability to transfer from natural language to genetic language, it indicates that the large model has developed a certain level of "understanding" of genetic sequences. This capability can provide novel tools and methods for biological research, expanding entirely new research and development approaches (Fig.8).

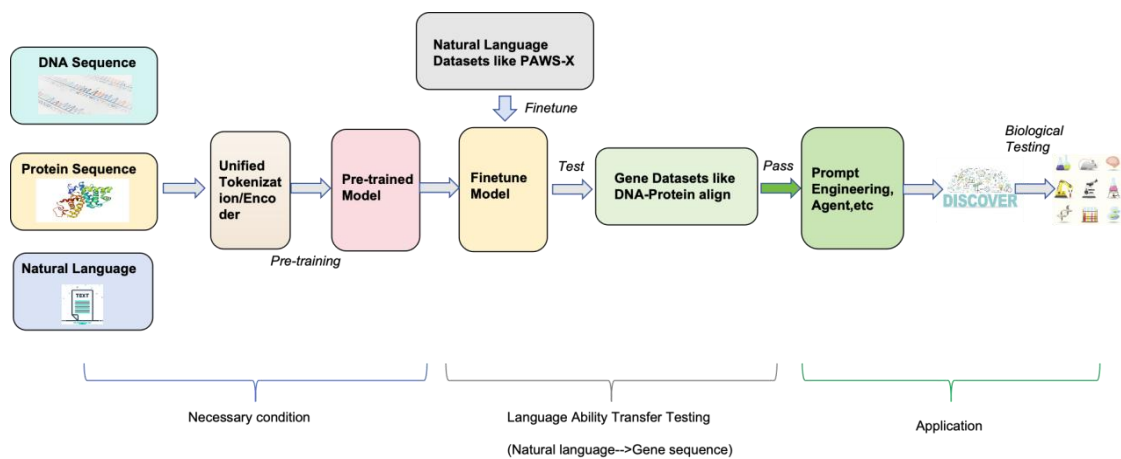

Fig.8. Expanding New Research Methods Based on the Transfer of Capabilities from Natural Language to Genetic Language.

The first step is establishing the fundamental prerequisites: pretraining a unified, multimodal large model using both natural language and biological sequences. Then, in the capability transfer validation phase, the model is fine-tuned with natural language datasets and tested on genetic language datasets. If the validation is successful, it indicates that the model has essentially mastered genetic language and can proceed to the application stage. In this stage, techniques such as prompt engineering can be used to discover new biological sequence patterns, which can then be validated through biological experiments.

Potential New Research Directions Enabled by Large Models in Understanding Natural and Genetic Languages:

1. Combining Prompt Engineering and Agents for Biological Research

    ○ Utilizing Prompt Engineering and Agents to enable large models to actively explore hidden patterns in genetic sequences. For example:

        ■ Designing structured prompts for DNA or protein sequences to predict potential functional regions, mutation effects, etc.

        ■ Training agents to autonomously search and optimize sequences to discover new functional genes or protein structures.

    ○ This approach can accelerate the discovery of new functional genes and optimize gene editing targets, such as those for CRISPR.

2. Expanding AI Applications in Biology Through Cross-Modal Learning

    ○ Integrating biological text data (e.g., research papers, patents) with experimental data to enhance the model's reasoning capabilities. For example:

        ■ Enabling large models to automatically read biomedical literature and generate new biological hypotheses by combining them with genetic sequence data.

        ■ Utilizing cross-modal AI to integrate imaging (e.g., protein structures), text, and sequence data for more precise disease prediction and personalized medicine analysis.

# 5 Conclusion

A large language model trained on both biological sequences and natural language text, after being fine-tuned on the Chinese-Korean similarity judgment dataset from PAWS-X, achieved a surprising test accuracy of 81% on the DNA-protein alignment dataset. In other words, an individual with the capabilities of a linguist who has observed a large number of DNA and protein sequences and learned how to judge the similarity between Chinese and Korean sentences could achieve an accuracy rate of over 81% in determining whether DNA and protein sequences match.

This study leverages the language transfer capabilities of large language models to achieve zero-shot prediction, uncovering the DNA-protein alignment coding rules.

The transfer of capabilities between natural languages can significantly enhance the ability of large models to serve low-resource languages. Similarly, transferring natural language capabilities to genetic language provides a novel perspective for studying biological languages, which remain largely unexplored. Through fine-tuning with datasets similar to PAWS-X and validation on DNA-protein alignment datasets, we have confirmed that large models can grasp the fundamental principles of genetic language.

On this basis, commonly used methods in large language models, such as prompt engineering and knowledge bases, can be applied. For example, through prompt engineering and agents, biological patterns can be discovered automatically. By combining experimental validation, research credibility can be enhanced. AI-powered data analysis pipelines can accelerate genomics and proteomics research. Cross-modal learning can further drive advancements in precision medicine and drug discovery. This direction not only contributes to fundamental biological research but also significantly enhances AI applications in life sciences, pharmaceuticals, and synthetic biology.

The code and data of this article are all open source (27).

# Data availability section

Sequence data that support the findings of this study have been deposited in huggingface.The fine-tuning/test data is located at https://huggingface.co/datasets/dnagpt/gene_lan_transfer, and the pre-training data and code is located at https://huggingface.co/dnagpt/central_dogma

# Supplementary Information

## 1 Data Process

### 1.1 Pretraining Data

The pretraining data mainly consists of DNA sequences, protein sequences, and text from seven different natural languages, with a maximum of 4GB per category.
The processed datasets are available at:
https://huggingface.co/dnagpt/central_dogma/tree/main/train_data

**DNA Sequences**
The DNA sequence data follows the dataset used in the DNABERT2 paper, which consists of randomly sampled sequences from the genomes of multiple model organisms. The original full 32GB dataset is stored at:
https://huggingface.co/dnagpt/llama-gene-train-data/tree/main/dna

**Protein Sequences**
The protein sequence data is sourced from the UniProt database (http://www.uniprot.org/downloads), including:

All data from UniProtKB/Swiss-Prot:
ftp://ftp.uniprot.org/pub/databases/uniprot/current_release/knowledgebase/complete/uniprot_sprot.fasta.gz

Selected data from UniProtKB/TrEMBL

ftp://ftp.uniprot.org/pub/databases/uniprot/current_release/knowledgebase/complete/uniprot_trembl.fasta.gz

Only the sequence portions from the FASTA files are extracted for use.

**Natural Language Data**

The natural language data is sourced from Wikipedia, with preprocessed datasets available on Hugging Face:

https://huggingface.co/datasets/legacy-datasets/wikipedia

https://huggingface.co/datasets/wikimedia/wikipedia

Both datasets are similar, differing mainly in update times. Either one can be selected, and only the title and text sequences are extracted for use.

## 1.2 Fine-tuning Data

The fine-tuning dataset used is PAWS-X, which is designed to evaluate cross-lingual transferability. The full dataset can be downloaded directly from Hugging Face:

https://huggingface.co/datasets/google-research-datasets/paws-x

Since PAWS-X contains similar sentence pairs within each language but does not provide cross-lingual sentence pairs, we reformatted it to create multi-language aligned data:

https://huggingface.co/datasets/dnagpt/paws-x-multi-pair

## 1.3 Testing Data

The testing dataset consists of DNA-protein sequence alignment pairs, available at:

https://huggingface.co/datasets/dnagpt/gene_lan_transfer

It includes:

1. A subset constructed by LucaOne, available at:

http://47.93.21.181/lucaone/DownstreamTasksDataset/dataset/CentralDogma/

LucaOne uses the NCBI-RefSeq database for DNA and protein matching, with a total of 8,533 positive samples. We performed a secondary extraction of the data using Biopython, removing the non-coding sequences of 100 base pairs from both ends, and then translated the remaining sequences into protein sequences. About 3,000 of these matched the target protein sequences, and only this portion was used as the positive examples in the dna_protein_pair dataset. For the remaining matching data, we needed to process the forward and reverse complement reading frames to search for ORFs. Therefore, we only selected the most clearly matched data as the positive examples for dna_protein_pair. The negative examples in LucaOne total 17,067, and a random subset was chosen as the negative examples for dna_protein_pair.

2. A custom-constructed dataset, where protein sequences are randomly sampled from UniProtKB/Swiss-Prot, and their corresponding DNA sequences are retrieved. For non-aligned pairs, the DNA sequence remains unchanged, while the protein sequence is randomly sampled from the protein database, ensuring its similarity to the correct protein sequence falls below a certain threshold.

The protein and DNA sequence pairs are also extracted from the NCBI database using Biopython. We also used Biopython for encoding analysis of the data. Out of the 8,000 positive examples, approximately 5,000 directly correspond with no gaps or mismatches. Around 2,700 require determining the reading frame before matching. There are about 200 samples with multiple shorter reading frames, which have relatively poor matches to the target protein sequence. This accounts for about 2.5% of the positive examples or 1.25% of the total dataset, which is considered normal error in ORF prediction. This does not significantly affect the testing results, so these samples have been retained.

The related data preprocessing scripts are available on Hugging Face:
https://huggingface.co/dnagpt/central_dogma/tree/main/get_data

## 2 Model List

The pre-trained and fine-tuned models used in this study have been uploaded to Hugging Face, as shown in Table.1.

Table.1 pre-trained model list

| Model name | huggingface path | description |
|---|---|---|
| gpt2_gene_multi_v1 | dnagpt/gpt2_gene_multi_v1 | GPT-2 model trained on DNA, protein, and seven natural languages. |
| gpt2_gene_multi_v2 | dnagpt/gpt2_gene_multi_v2 | GPT-2 model trained on DNA, protein, and seven natural languages. |
| gene_eng_zh | dnagpt/gene_eng_zh | GPT-2 model trained on DNA, protein, English, and Chinese. |
| gene_eng_zh_de_es | dnagpt/gene_eng_zh_de_es | GPT-2 model trained on DNA, protein, English, Chinese, German, and Spanish. |
| gpt2_gene_multi_v1_ft | dnagpt/gpt2_gene_multi_v1_ft | Model fine-tuned on English data from PAWS-X. |
| gpt2_gene_multi_v2_ft | dnagpt/gpt2_gene_multi_v2_ft | Model fine-tuned on English data from PAWS-X. |
| gene_eng_gpt2_v1 | dnagpt/gene_eng_gpt2_v1 | GPT-2 model trained on DNA, protein, and English text |
| gpt2 | gpt2 | Standard GPT-2 model |

## 3 Effect of Fine-Tuning on Different Language

**Table.2. Effect of Fine-Tuning on Different Language Datasets, en-other language pair. Accuracy result.**

| finetune data pair en--> | test dna_protein_pair | test dna_protein_pair_rand |
| --- | --- | --- |
| en | - | - |
| Fr | 0.69 | 0.78 |
| De | 0.70 | 0.75 |
| Es | 0.67 | 0.73 |
| Zh | 0.70 | 0.72 |
| Ja | 0.65 | 0.73 |
| Ko | 0.68 | 0.76 |

Table.3. Effect of Fine-Tuning on Different Language Datasets, fr-other language pair

| finetune data pair fr--> | test dna_protein_pair | test dna_protein_pair_rand |
| --- | --- | --- |
| en | 0.72 | 0.72 |
| Fr | - | - |
| De | 0.7 | 0.74 |
| Es | 0.70 | 0.76 |
| Zh | 0.70 | 0.69 |
| Ja | 0.68 | 0.69 |
| Ko | 0.70 | 0.76 |

Table.4. Effect of Fine-Tuning on Different Language Datasets, de-other language pair

| finetune data pair de--> | test dna_protein_pair | test dna_protein_pair_rand |
| --- | --- | --- |
| en | 0.71 | 0.73 |
| Fr | 0.71 | 0.73 |
| De | - | - |
| Es | 0.72 | 0.73 |
| Zh | 0.74 | 0.76 |
| Ja | 0.65 | 0.72 |
| Ko | 0.66 | 0.71 |

Table.5. Effect of Fine-Tuning on Different Language Datasets, es-other language pair . Accuracy result.

| finetune data pair es--> | test dna_protein_pair | test dna_protein_pair_rand |
|---|---|---|
| en | 0.73 | 0.72 |
| Fr | 0.67 | 0.69 |
| De | 0.71 | 0.79 |
| Es | - | - |
| Zh | 0.70 | 0.74 |
| Ja | 0.66 | 0.71 |
| Ko | 0.70 | 0.77 |

Table.6. Effect of Fine-Tuning on Different Language Datasets, zh-other language pair. Accuracy result.

| finetune data pair zh--> | test dna_protein_pair | test dna_protein_pair_rand |
|---|---|---|
| en | 0.71 | 0.68 |
| Fr | 0.72 | 0.67 |
| De | 0.71 | 0.71 |
| Es | 0.7 | 0.71 |
| Zh | - | - |
| Ja | 0.72 | 0.77 |
| Ko | 0.68 | 0.81 |

Table.7. Effect of Fine-Tuning on Different Language Datasets, ja-other language pair. Accuracy result.

| finetune data pair ja--> | test dna_protein_pair | test dna_protein_pair_rand |
|---|---|---|
| en | 0.68 | 0.66 |
| Fr | 0.69 | 0.72 |
| De | 0.72 | 0.75 |
| Es | 0.68 | 0.79 |
| Zh | 0.69 | 0.75 |
| Ja | - | - |
| Ko | 0.67 | 0.74 |

Table.8. Effect of Fine-Tuning on Different Language Datasets, ko-other language pair. **Accuracy result.**

| finetune data pair ko--> | test dna_protein_pair | test dna_protein_pair_rand |
|---|---|---|
| en | 0.72 | 0.72 |
| Fr | 0.73 | 0.72 |
| De | 0.70 | 0.70 |
| Es | 0.71 | 0.72 |
| Zh | 0.72 | 0.68 |
| Ja | 0.66 | 0.71 |
| Ko | - | - |